\documentclass[jkps,preprint,fleqn,showpacs,showkeys]{revtex4}
\usepackage{graphicx}
\usepackage{amssymb}
\usepackage{amsmath}
\usepackage{bm}
\begin{document}
\setcounter{page}{0}
\title[]{Dirac Coupled Channel Analyses of the high-lying excited states at $^{22}$Ne(p,p$'$)$^{22}$Ne}
\author{ Sugie \surname{Shim}}
\email{shim@kongju.ac.kr}
\thanks{Fax: +82-41-850-8489}
\author{and Moon-Won \surname{Kim} }
\affiliation{Department of Physics, Kongju National University, Gongju 314-701}

\date[]{Received 2015}

\begin{abstract}
Dirac phenomenological coupled channel analyses are performed using an
optical potential model for the high-lying excited vibrational states at 800 MeV unpolarized proton inelastic scatterings from $^{22}$Ne nucleus. Lorentz-covariant
scalar and time-like vector potentials are used as direct optical potentials and the first-order vibrational
collective model is used for the transition optical potentials to
describe the high-lying excited vibrational collective states. The complicated
Dirac coupled channel equations are solved phenomenologically using a sequential iteration method by varying the optical potential and the deformation parameters. Relativistic Dirac coupled channel calculations are able to describe the high-lying excited states of the vibrational bands in $^{22}$Ne clearly better than the nonrelativistic coupled channel calculations.
The channel-coupling effects of the multistep process for the excited states of the vibrational
bands are investigated. The deformation parameters obtained from the Dirac phenomenological calculations for the high-lying vibrational excited states in $^{22}$Ne are found to agree well with those obtained from the nonrelativistic calculations using the same Woods-Saxon potential shape.
\end{abstract}



\maketitle

\section{Introduction}

Relativistic treatment of nuclear reactions based on use of the Dirac equation have proven to be very successful, in particular for the description of elastic and inelastic nucleon-nucleus scatterings\cite{1,2,3,4,5,6}.
Considerable improvements have been obtained in the Dirac coupled channel calculations for the intermediate energy proton inelastic scatterings from spherically symmetric nuclei and from a few deformed nuclei compared to the conventional nonrelativistic calculations based on the Schr\"{o}dinger equation\cite{4,5,6,7,8,9,10,11}. We should note that one of the merits of the relativistic approach based on the Dirac equation instead of a nonrelativistic approach based on the Schr\"{o}dinger equation is that the spin-orbit potential appears naturally in the Dirac approach when the Dirac equation is reduced to the Schr\"{o}dinger-like second-order differential equation, whereas the spin-orbit potential has to be inserted by hand in the nonrelativistic Schr\"{o}dinger approach.

In this work, we performed Dirac coupled channel analyses for the high-lying excited vibrational states in an s-d shell nucleus $^{22}$Ne at intermediate energy proton inelastic scatterings. An optical potential model called S-V model is used and only scalar and time-like vector potentials are considered for the direct optical potentials. The Woods-Saxon shape is used for the geometry of the direct optical potentials.
In order to accommodate the collective  motion of the excited deformed nucleus considering the high-lying excited states that belong to the vibrational
bands in $^{22}$Ne, we use the first-order vibrational collective model to obtain the transition optical potentials. 2$^-$ and 3$^-$ excited states of the $K^\pi = 2^- $ octupole vibrational band, and the second 2$^+$ and the second 4$^+$ excited states those are assumed to be the members of the $K^\pi = 2^+ $ gamma vibrational band are considered in the calculation.
The complicated Dirac coupled channel equations are solved using a computer program called ECIS\cite{12}, where the Dirac optical potential and deformation parameters are determined phenomenologically using a sequential iteration method. The channel-coupling effects of the multistep process for the high-lying excited states of the vibrational
bands are investigated.  The results of the Dirac coupled channel calculations for the 800MeV proton inelastic scatterings from $^{22}$Ne are analyzed and compared with the experimental data and with the results obtained from the nonrelativistic approaches.

\section{Theory and Results}

Dirac coupled channel calculations are performed phenomenologically for the high-lying excited states that possibly belong to the 2$^-$ or the 2$^+$ vibrational
bands at the 800-MeV unpolarized proton inelastic scatterings from $^{22}$Ne by using an optical potential model and the first-order vibrational collective model.
Because $^{22}$Ne is one of the spin-0 nuclei, only scalar, time-like vector, and tensor
optical potentials can survive\cite{13,14}; hence, the relevant Dirac equation for the elastic scattering from the nucleus is given as
\begin{equation}
[\alpha \cdot p + \beta ( m + U_S ) - ( E - U_0- V_c )
 + i \alpha \cdot  \hat{r} \beta U_T ] \Psi(r) = 0.
\label{e1}
\end{equation}
Here, $U_S$ is a scalar potential, $U_0 $ is a time-like vector potential,  $U_T$ is a tensor potential, and $V_c $ is the Coulomb potential.
However, it is also true that pseudo-scalar and
axial-vector potentials may be present in the equation when we consider inelastic
scattering, depending on the  model assumed. In this work, we assume that appropriate transition potentials can be obtained by deforming the direct potentials that describe the elastic
channel reasonably well\cite{15}.
Transition potentials are obtained by assuming that they are proportional to the first-order derivatives of the diagonal potentials. The scalar and the time-like vector potentials are used as direct potentials in the calculation. Even though tensor potentials are always present due to the interaction of the anomalous magnetic moment of the projectile with the charge distribution of the target, they have been found to be always very small compared to the scalar or the vector potentials\cite{4}. Hence, they are neglected in this calculation. The evidence that the large scalar and vector fields of Dirac phenomenology may be related to quark degrees of freedom in the nucleon can be found in the work of Cohen, Furnstahl, and Griegel\cite{16}.
In the vibrational model of ECIS, the deformation of the nuclear surface is written using the Legendre polynomial expansion method as
\begin{equation}
R(\theta, \phi ) = R_0 ( 1+ \sum_{\lambda \mu } \beta_\lambda Y^* _{\lambda \mu } (\theta, \phi ) ),
\label{e2}
\end{equation}
with $R_0$ being the radius at equilibrium, $\beta$ a deformation parameter and $\lambda$ the multipolarity.
  The transition potentials are given by
\begin{equation}
U_i ^\lambda = \sum_\mu \frac{\beta^i _\lambda R_i }{(2\lambda +1)^{1/2}} \frac{dU_i (r)}{dr } Y^* _{\lambda \mu} (\Omega),
\label{e3}
\end{equation}
where the subscript $i$  refers to the real and the imaginary scalar or vector potential and $R$ is the radius parameter of the Woods-Saxon shape. The real and the imaginary deformation parameters are taken to be equal for a given  potential type so that two deformation parameters, $\beta_S$ and $\beta_V$, are determined for each excited state.

The experimental data for the differential cross sections are obtained from Ref. 17 for the 800 MeV unpolarized proton inelastic scatterings from $^{22}$Ne nucleus. The high-lying excited states of the 2$^-$ octupole vibrational band, the $2^-$(5.15 MeV) and $3^-$(5.91 MeV) states, and possibly those of 2$^+$ gamma vibrational band, the second $2^+$(4.46 MeV) and the second $4^+$(5.52 MeV) states, are considered and assumed to be collective vibrational states in the calculation.
The 5$^-$ state that could belong to the 2$^-$ octupole vibrational band is not considered in the calculation because the experimental data for the state do not exist\cite{17}.

The 12 parameters of the diagonal scalar and vector potentials in the Woods-Saxon shapes are determined phenomenologically by fitting the experimental elastic scattering differential cross section data.
The calculated optical potential parameters for the 800 MeV proton elastic scatterings from $^{22}$Ne are shown in Ref. 18.
Firstly, we perform six-parameter searches by including one excited state, the $2^-$ state or the $3^-$ state of the 2$^-$ vibrational band, or the second $2^+$ state ($2_2^+$) state or the second $4^+$ state ($4_2^+$) state of the 2$^+$ vibrational band, in addition to the ground state, starting from the obtained 12 parameters for the  direct optical potentials. Here, six parameters determine the four potential strengths, the scalar real and imaginary potential strengths and the vector real and imaginary potential strengths, keeping the potential geometry unchanged, and the two deformation parameters $\beta_S$ and $\beta_V$ of the included excited state. Here, the four optical potential strengths obtained by fitting to the elastic scattering data vary because the channel coupling of the excited states to the ground state should be included in the inelastic scattering calculation.
The Dirac coupled channel equations are solved to obtain the best fitting parameters to the experimental data by using the minimum $\chi^2$ method numerically.
Next, eight-parameter searches are performed by including the $2^-$ and the $3^-$ excited states of the 2$^-$ vibrational band, or $2_2^+$ state and $4_2^+$ state of the 2$^+$ vibrational band in addition to the ground state.
Because we found the best fit parameters when we include all three states, we performed two-parameter searches($\beta_S$ and $\beta_V$) for the cases where only one excited state is coupled in order to investigate the pure channel coupling effect among the excited states. Here, two parameters are $\beta_S$ and $\beta_V$, fixing the potential strengthes using the values obtained where the ground, the $2^-$ and the $3^-$ excited states of the 2$^-$ vibrational band, or the ground, the $2_2^+$ and the $4_2^+$ states of the 2$^+$ vibrational band are coupled in the calculation.

The results of the Dirac coupled channel calculations for the ground state are given in Fig. 1, when the excited states of the $2^-$ vibrational band are coupled. In the figures, `cpd' means `coupled'.
We show the results of the 2-parameter and 8-parameter searches in the figures and tables in order to investigate the pure coupled channel effects. Figure 1 shows that most of the calculations reproduce the elastic experimental data quite well except for the case where only the $2^-$ state is coupled. For the case where only the $2^-$ state coupled, even the ground state is not well described. This seems to be due to the fact that the inelastic excitation of the $2^-$ state itself is dominated by the sequential two-step process via the $3^-$ state rather than the direct process from the elastic channel to the $2^-$ inelastic channel\cite{17}. Similar discrepancy has been found in the calculation for the proton scattering from  $^{20}$Ne\cite{19}. The calculated observables for the $2^-$ state are shown in Fig. 2. The dashed and the solid lines represent the results of the calculations where the ground and the $2^-$ states are coupled, and where the ground, the $2^-$ and the $3^-$ states are coupled, respectively. Even though still the agreement with the data is not so good, missing the minimum positions of the diffraction pattern, the agreements with the $2^-$ data are improved noticeably by adding the coupling with the $3^-$ state.  Thus, including the multi-step transition process via the channel coupling with the $3^-$ state seems important for describing the excitation of the $2^-$ state\cite{9,17}. This feature is also found previously for the proton inelastic scatterings from $^{20}$Ne.\cite{19} This feature can be explained by that the coupling between the ground state and $K^{\pi}=2^-$ band has the structure {$\alpha_{32} (Y_{32} + Y_{3-2} ) $\cite{17}. Hence the $2^-$ state can be excited only by transitions which involve at least two steps, not from the ground state directly.

\begin{figure}
\includegraphics[width=10.0cm]{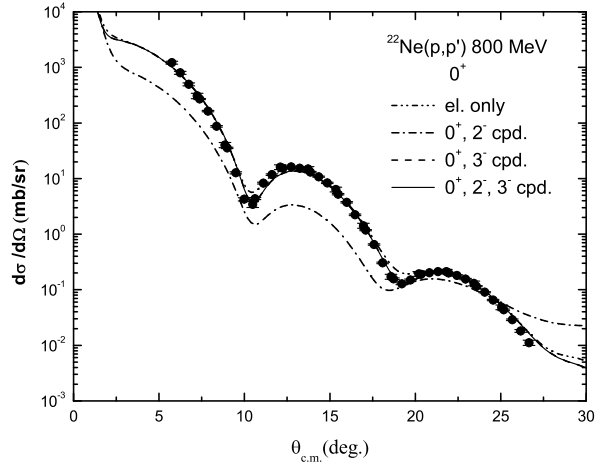}
\caption{Differential cross section of the 800 MeV p +  $^{22}$Ne elastic scattering. The dash-dot-dot, the dash-dot, the dashed, and the solid lines represent the results of Dirac phenomenological calculations where only the ground state is considered, where the ground and the $2^-$ states are coupled, where the ground and the $3^-$ states are coupled, and where the ground, the $2^-$  and the $3^-$ states of the 2$^-$ vibrational band are coupled, respectively. }
\label{fig1}
\end{figure}

\begin{figure}
\includegraphics[width=10.0cm]{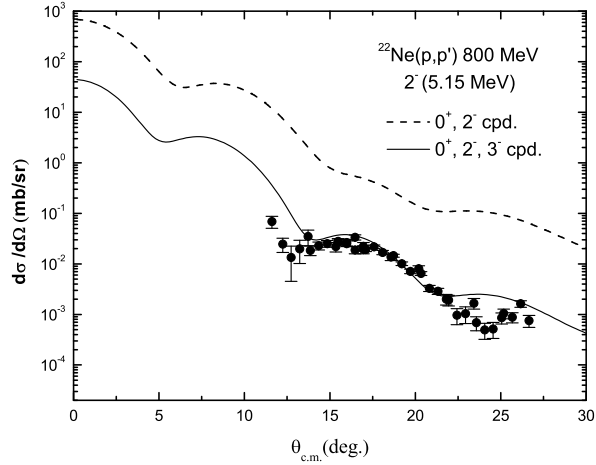}
\caption{Differential  cross section of the $2^- $ state for 800 MeV p +  $^{22}$Ne inelastic scattering. The dashed and the solid lines represent the results of the Dirac coupled channel calculations where the ground and the $2^-$ states are coupled, and where the ground, the $2^-$ and the $3^-$ states of the 2$^-$ vibrational band are coupled, respectively.}
\label{fig2}
\end{figure}

It is true that there exist the low-lying excited states that belong to the ground state rotational band (GSRB) and they can couple with the high-lying excited states. However, it is not likely the excited states of the 2$^-$ vibrational band to couple with the excited states of GSRB due to the parity violation.
Figure 3 shows the calculated results for the excitation of the $3^-$ state. The agreements with experimental data for the $3^-$ state turn out to be quite good for the case where the $2^-$ and the $3^-$ states are included, showing clearly better agreements with the experimental data compared to the results obtained from the nonrelativistic calculations\cite{17}. Even for the case where only the $3^-$ state is coupled to the ground state, the agreements with the experimental data turn out to be pretty good, indicating that excitation by a direct transition from the ground state is dominant for this state.

\begin{figure}
\includegraphics[width=10.0cm]{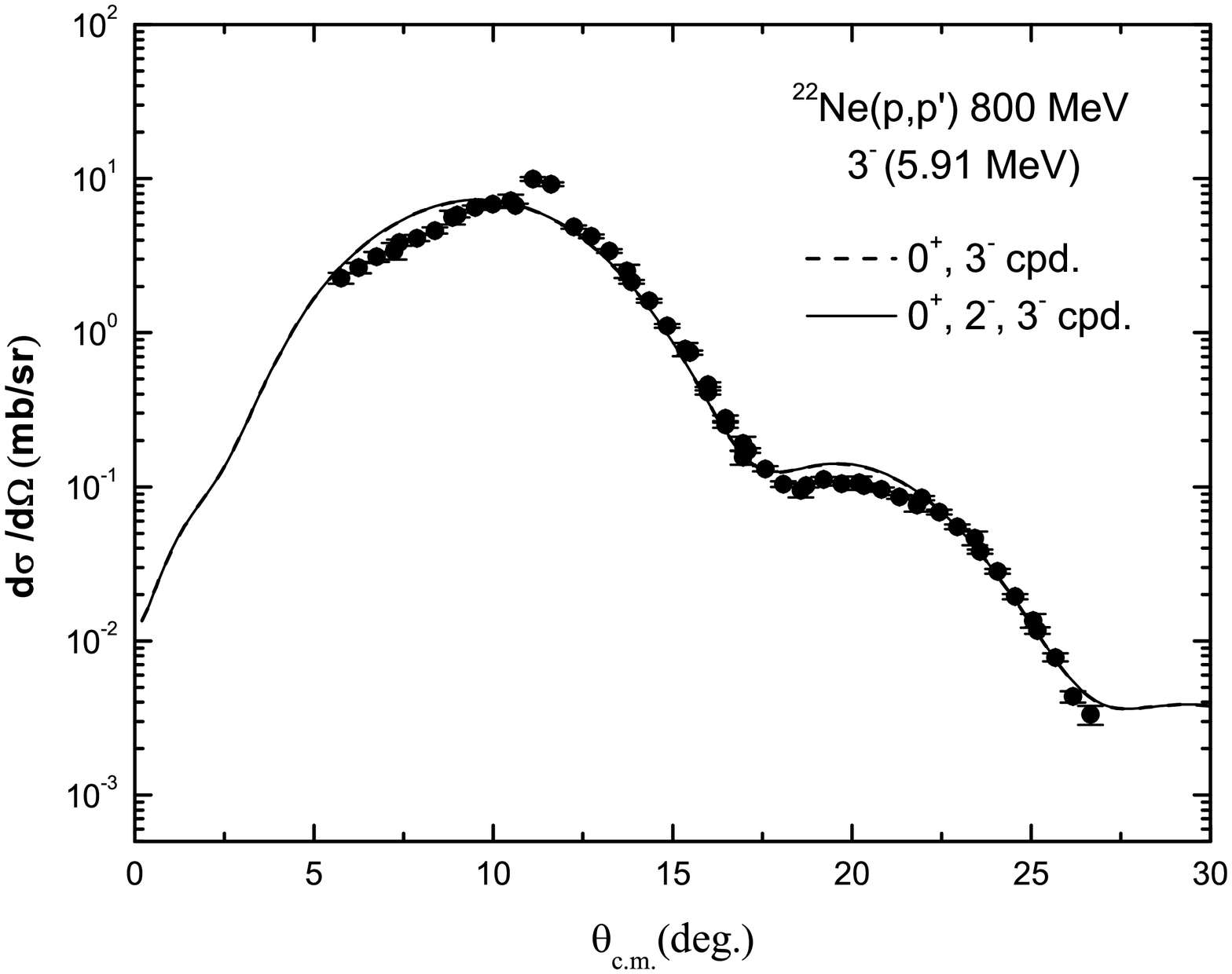}
\caption{Differential  cross section of the $3^- $ state for 800 MeV p +  $^{22}$Ne inelastic scattering. The dashed and the solid lines represent the results of Dirac coupled channel calculations where the ground and the $3^-$ states are coupled, and where the ground, the $2^-$ and the $3^-$ states of the 2$^-$ vibrational band are coupled, respectively.}
\label{fig3}
\end{figure}

\begin{table}
\caption{ Comparison of the deformation  parameters for the excited states of the 2$^-$ vibrational band in $^{22}$Ne for 800 MeV proton inelastic scatterings with those obtained by using  the nonrelativistic calculations$^{17}$. Potential strengths are ordered as scalar real, scalar imaginary, vector real, and vector imaginary, downward from the top.}
\begin{ruledtabular}
\begin{tabular}{ccccccccccc} 
        
         &   E   & $\chi_{e}^{2}$/N & $\chi_{i}^{2}$/N & $\beta_{S}$ & $\beta_{V}$ & $\beta_{NR}$ & Potential ~ \\
         & (MeV) &      &         &            &            &             & (MeV) ~ \\ \hline
 Elastic &      & 22.5 &     &      &      &      & -253.5  ~ \\
 only    &      &     &     &      &      &      &  192.3  ~ \\
         &      &     &     &      &      &      &  107.0  ~ \\
         &      &     &     &      &      &      & -102.6 ~ \\ \hline
 2$^-$   & 5.15 & 1223 & 1528 & 0.177 & 0.976 &      &   -253.5  ~ \\
  cpd.   &      &     &     &      &      &      &    192.3  ~ \\
         &      &     &     &      &      &      &   107.0  ~ \\
         &      &     &     &      &      &      &  -102.6  ~ \\ \hline
 3$^-$   & 5.91 & 15.9 & 31.5 & 0.286 & 0.251 &      &   -253.5  ~ \\
  cpd.   &      &     &     &      &      &      &    192.3 ~ \\
         &      &     &     &      &      &      &   107.0  ~ \\
         &      &     &     &      &      &      &  -102.6  ~ \\ \hline
 2$^-$   & 5.15 & 14.9 & 9.95 & -0.053 & -0.042 &      &  -253.5  ~ \\
 3$^-$   & 5.91 &     & 31.4 & 0.287 & 0.252 &  0.243    &   192.3  ~ \\
  cpd.   &      &     &     &      &      &      &   107.0  ~ \\
         &      &     &     &      &      &      &  -102.6  ~ \\
\end{tabular}
\end{ruledtabular}
\label{table1}
\end{table}

In Table 1, we show the deformation parameters for the excited states of the 2$^-$ vibrational band in $^{22}$Ne obtained in the 2-parameter searches for only one excited state coupled cases and in the 8-parameter search for the two excited states coupled case, and compare them with those obtained by using the nonrelativistic coupled channel calculation. $\chi_{e}^{2}$/N and $\chi_{i}^{2}$/N are reduced chi-square values for the elastic and the inelastic
cross section fitting results, respectively. Here, N is the number of experimental cross section data for each state. The results of Dirac phenomenological calculation are seen to agree pretty well with the result of the nonrelativistic calculation for the $3^-$ state\cite{17}. $\beta_{V}$ shows almost the same value with the deformation parameter of the nonrelativistic calculation, while $\beta_{S}$ shows slightly smaller value. Also, the changes in the potential strengths and the values of $\chi^{2}$/Ns are given in Table 1 for the eight-parameter searches. We note that when the 2$^-$ and 3$^-$ states are coupled, the best fit results for the excited states are obtained. By adding the 3$^-$ state, the value of $\chi^{2}$/N for the 2$^-$ state is reduced to less than 1/150 of the value of $\chi^{2}$/N obtained when only the 2$^-$ state is coupled to the ground state, confirming that the multi-step excitation process via channel coupling with the 3$^-$ state might be important for the excitation of the 2$^-$ state. However, the value of $\chi^{2}$/N for the 3$^-$ state is almost not changed, confirming that a direct excitation from the ground state is dominant for the 3$^-$ state. These features are the same with those found previously for the proton inelastic scatterings from $^{20}$Ne\cite{19}. The deformation parameter $\beta_{V}$ is observed to be larger than $\beta_{S}$ for the 3$^-$ state, as observed for the scatterings from other deformed nuclei in the Dirac coupled channel calculations\cite{11}.

\begin{figure}
\includegraphics[width=10.0cm]{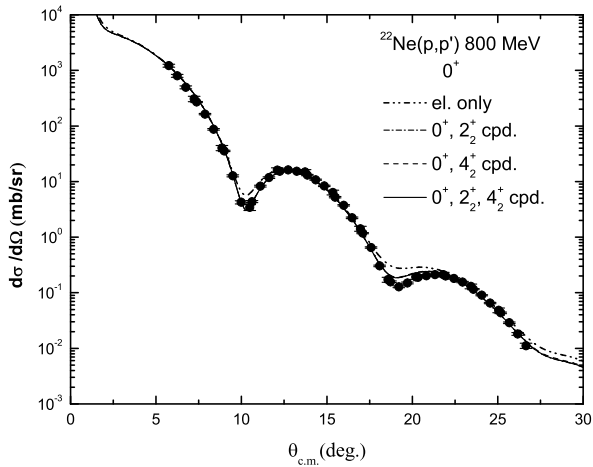}
\caption{Differential cross section of the 800 MeV p +  $^{22}$Ne elastic scattering. The dash-dot-dot, the dash-dot, the dashed, and the solid lines represent the results of Dirac phenomenological calculations where only the ground state is considered, where the ground and the $2_2^+$ states are coupled, where the ground and the $4_2^+$ states are coupled, and where the ground, the $2_2^+$  and the $4_2^+$ states of the 2$^+$ vibrational band are coupled, respectively. }
\label{fig1}
\end{figure}

 In Fig. 4 we show the results of the Dirac coupled channel calculations for the ground state when the excited states of the $2^+$ vibrational band are coupled. Figure 4 shows that all of the calculations reproduce the elastic experimental data quite well. It is shown that the channel coupling effect with the  $2_2^+$  and the $4_2^+$ states of the 2$^+$ vibrational band moved the theoretical line slightly downward to the experimental data at the first and second minima of the diffraction pattern, improving the agreement with the experimental data.
In Fig. 5, we show the results of the Dirac coupled channel calculations for the $2_2^+$ state, which is assumed to be the lowest-lying excited state of the 2$^+$ gamma vibrational band, and the agreements with the experimental data turn out to be not so good.
The dashed and the solid lines represent the results of the calculations where the ground and the $2_2^+$ states are coupled, and where the ground, the $2_2^+$ and the $4_2^+$ states are coupled, respectively. The agreements with the $2_2^+$ differential cross section data are not improved noticeably even when the channel coupling with the $4_2^+$ state is added in the calculation.
In order to improve the agreement with the $2_2^+$ data, even the channel coupling effect with the third $4^+$ state (6.34 MeV)\cite{17} is investigated. But still the agreements with the experimental data are not improved noticeably.
This could be explained by the fact that the assignment of the $2_2^+$ state to the $K^\pi = 2^+ $ band is not certain, as the data can be explained as belonging to a $K^\pi = 0^+ $ band\cite{17}, so it might be necessary to include couplings with other unmeasured excited states nearby to describe this state well. Also, in this case the coupling with the excited states of GSRB could play important role, but they are not included to investigate the channel coupling between the excited states that assumed to belong to the 2$^+$ gamma vibrational band  in the calculation. We plan to investigate the channel coupling effect between the excites states of the GSRB and those of the high-lying vibrational band in the near future.

\begin{figure}
\includegraphics[width=10.0cm]{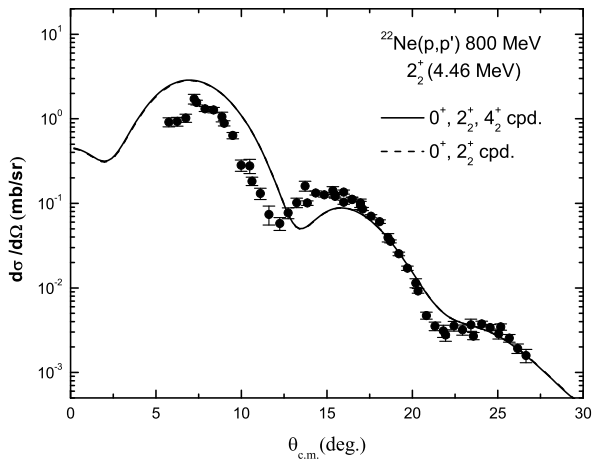}
\caption{Differential  cross section of the $2_2^+ $ state for 800 MeV p +  $^{22}$Ne inelastic scattering. The dashed and the solid lines represent the results of Dirac coupled channel calculations where the ground and the $2_2^+$ states are coupled, and where the ground, the $2_2^+$ and the $4_2^+$ states are coupled, respectively.}
\label{fig4}
\end{figure}

\begin{figure}
\includegraphics[width=10.0cm]{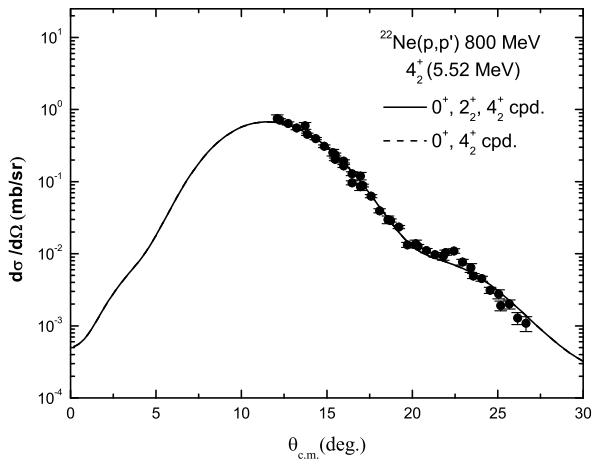}
\caption{Differential  cross section of the $4_2^+ $ state for 800 MeV p +  $^{22}$Ne inelastic scattering. The dashed and the solid lines represent the results of Dirac coupled channel calculations where the ground and the $4^+$ states are coupled, and where the ground, the $2_2^+$ and the $4_2^+$ states are coupled, respectively.}
\label{fig5}
\end{figure}

Figure 6 shows the calculated results for the excitation of the $4_2^+$ state. Even for the case where only the $4_2^+$ state is coupled to the ground state, the agreements with the experimental data turn out to be pretty good  as shown by the dashed lines, and they are not changed noticeably by adding the coupling with the 2$_2^+$ state, as shown by solid lines.
In Table 2, we show the deformation parameters of the Lorentz covariant scalar and vector optical potentials for the 2$_2^+$ and 4$_2^+$ excited states in $^{22}$Ne and compare them with those obtained by using the nonrelativistic coupled channel calculations. The results of Dirac phenomenological calculations are seen to agree pretty well with the results of the nonrelativistic calculations. Again, $\beta_{V}$ shows better agreement with the deformation parameter of the nonrelativistic calculation, while $\beta_{S}$ shows slightly smaller value. By adding the coupling with the 4$_2^+$ state, the value of $\chi^{2}$/N for the 2$_2^+$ state is reduced just a little. Also, by adding the coupling with the 2$_2^+$ state, the value of $\chi^{2}$/N for the 4$_2^+$ state is almost not changed compared with the value of $\chi^{2}$/N obtained when only the 4$_2^+$ state is coupled to the ground state, indicating that excitation by a direct transition from the ground state is dominant for the 4$_2^+$ state in $^{22}$Ne.  It is shown that
 an $\alpha_{42} (Y_{42} + Y_{4-2} ) $ coupling which permits a strong direct transition from the ground state to the $4^+$ member of the $\gamma$ vibrational band is required for the satisfactory coupled channel calculations in the analyses of $K^{\pi}=2^+$ bands in $^{24,26}$Mg, by Blanpied ${\it et\ al.}$ \cite{17}.
Again, the deformation parameter $\beta_{V}$ is observed to be larger than $\beta_{S}$ for the 4$_2^+$ state, as observed at the scatterings from other deformed nuclei such as $^{24}$Mg, $^{26}$Mg, $^{154}$Sm and $^{176}$Yb\cite{11}.

\begin{table}
\caption{ Comparison of the deformation  parameters for the 2$_2^+$ and 4$_2^+$ excited states in $^{22}$Ne for 800 MeV proton inelastic scatterings with those obtained by using the nonrelativistic calculations.$^{17}$ Potential strengths are ordered as scalar real, scalar imaginary, vector real, and vector imaginary, downward from the top.}
\begin{ruledtabular}
\begin{tabular}{ccccccccccc}
        
         &   E   & $\chi_{e}^{2}$/N & $\chi_{i}^{2}$/N & $\beta_{S}$ & $\beta_{V}$ & $\beta_{NR}$ & Potential ~ \\
         & (MeV) &      &         &            &            &             & (MeV) ~ \\ \hline
 Elastic &      & 44.1 &     &      &      &      & -251.9  ~ \\
 only    &      &     &     &      &      &      &  151.0  ~ \\
         &      &     &     &      &      &      &   108.5  ~ \\
         &      &     &     &      &      &      & -98.82 ~ \\ \hline
 2$_2^+$   & 4.46 & 11.74 & 38.0 & 0.074 & 0.087 &      & -251.9  ~ \\
  cpd.   &      &     &     &      &      &      &   151.0  ~ \\
         &      &     &     &      &      &      &   108.5  ~ \\
         &      &     &     &      &      &      &  -98.82  ~ \\ \hline
 4$_2^+$   & 5.52 & 12.3 & 4.31 & 0.067 & 0.078 &      & -251.9  ~ \\
  cpd.   &      &     &     &      &      &      &   151.0  ~ \\
         &      &     &     &      &      &      &   108.5  ~ \\
         &      &     &     &      &      &      &  -98.82  ~ \\ \hline
 2$_2^+$   & 4.46 & 11.5 & 37.9 & 0.072 & 0.086 & 0.093     & -251.9  ~ \\
 4$_2^+$   & 5.52 &     & 4.32 & 0.067 & 0.078 &   0.100   &   151.0  ~ \\
  cpd.   &      &     &     &      &      &      &   108.5  ~ \\
         &      &     &     &      &      &      &  -98.82  ~ \\
\end{tabular}
\end{ruledtabular}
\label{table2}
\end{table}

\section{Conclusions}

Dirac phenomenological coupled channel analyses are performed using an
optical potential model for the high-lying excited vibrational states at 800 MeV unpolarized proton inelastic scatterings from $^{22}$Ne nucleus. The channel-coupling effects of the multistep process for the excited states of the vibrational bands are investigated.
Relativistic Dirac coupled channel calculations are able to describe the high-lying excited states for the proton inelastic scatterings from an s-d shell nucleus $^{22}$Ne clearly better than the nonrelativistic coupled channel calculations.
 The first-order vibrational collective models are used to describe the high-lying excited vibrational states in the nucleus. The deformation parameters of the Lorentz covariant scalar and vector optical potentials obtained from the Dirac phenomenological calculations for the high-lying vibrational excited states in $^{22}$Ne are found to agree well with those obtained from the nonrelativistic calculations using the same Woods-Saxon potential shape.
 A pure direct transition from the ground state is shown to be dominant for the excitation of the 3$^-$ state and multi-step excitation via channel coupling with the 3$^-$ state is important for the excitation of the $2^-$ state in the 2$^-$ octupole vibrational band in $^{22}$Ne, which are the same features found previously at the inelastic scatterings from $^{20}$Ne.
 For the excitation of the 4$_2^+$ state in $^{22}$Ne, the direct transition from the ground state is shown to be dominant.
 The deformation parameter $\beta_{V}$ is observed to be larger than $\beta_{S}$ for the excited states considered in the calculation, as previously observed at the scatterings from other deformed nuclei such as $^{24}$Mg, $^{26}$Mg, $^{154}$Sm and $^{176}$Yb in the Dirac coupled channel calculations.


\begin{references}

\bibitem{1} L. G. Arnold, B. C. Clark, R. L. Mercer, and  P. Swandt, Phys. Rev. C {\bf 23}, 1949 (1981).
\bibitem{2} B. C. Clark, R. L. Mercer, and  P. Swandt, Phys. Lett. {\bf 122B}, 211 (1983).
\bibitem{3} S. Hama, B. C. Clark, R. E. Kozack, S. Shim, E. D. Cooper, R. L. Mercer, and B. D. Serot, Phys. Rev. C {\bf 37} 1111, (1988).
\bibitem{4} S. Shim, Ph.D. dissertation, The Ohio State University 1989; L. Kurth, B. C. Clark, E. D. Cooper, S. Hama, S. Shim, R. L. Mercer, L. Ray, and G. W. Hoffmann, Phys.  Rev. C {\bf 49}, 2086 (1994).
\bibitem{5} S. Shim, B. C. Clark, E. D. Cooper, S. Hama, R. L. Mercer, L. Ray, J. Raynal, and H. S. Sherif, Phys. Rev. C {\bf 42}, 1592 (1990).
\bibitem{6} R. de Swiniarski, D. L. Pham, and J. Raynal, Z. Phys. {\it Z. Phys. A-Hadrons and Nuclei} {\bf 343}, 179 (1992).
\bibitem{7} D. L. Pham and R. de Swiniarski, Nuovo Cimento A {\bf 107}, 1405 (1994).
\bibitem{8} J. J. Kelly, Phys. Rev. C {\bf71}, 064610 (2005).
\bibitem{9} S. Shim, M. W. Kim, B. C. Clark, and L. Kurth Kerr, Phys. Rev. C {\bf 59}, 317 (1999).
\bibitem{10} S. Shim, Shin-Ho Ryu and Min-Soo Kim, J. Korean. Phys. Soc. {\bf 51}, 271 (2007).; S. Shim, Shin-Ho Ryu and Min-Soo Kim, J. Korean. Phys. Soc. {\bf 53}, 1146 (2008).
\bibitem{11} S. Shim and M. W. Kim, Int. J. Mod. Phys. E {\bf 21}, 1250098 (2012).
\bibitem{12} J. Raynal, {\it Computing as a Language of Physics}, ICTP International Seminar Course, 281 (IAEA, Italy, 1972); J. Raynal, {\it Notes on ECIS94}, Note CEA-N-2772, 1994.
\bibitem{13} C. J. Horowitz and B. D. Serot, Nucl. Phys. A {\bf 368}, 503 (1981).
\bibitem{14} R. J.  Furnstahl, C. E.  Price, and G.  E. Walker, Phys.  Rev. C  {\bf 36}, 2590 (1987).
\bibitem{15} L. Ray and G. W. Hoffmann, Phys. Rev. C {\bf 31}, 538 (1986).
\bibitem{16} T. D. Cohen, R. J. Furnstahl, and D. K. Griegel, Phys. Rev. Lett. {\bf 67}, 961 (1991).
\bibitem{17} G. S. Blanpied, B. G. Ritchie, M. L. Barlett, R. W. Fergerson, G. W. Hoffmann, J. A. McGill, B. H. Wildenthal, Phys. Rev. C {\bf 38}, 2180 (1988); Experimental Nuclear Reaction Data(EXFOR), www-nds.iaea.org.
\bibitem{18} Moon-Won Kim and Sugie Shim, J. Korean. Phys. Soc. {\bf 66}, 850 (2015).
\bibitem{19} S. Shim, J. Korean. Phys. Soc. {\bf 65}, 1179 (2014).
\end{references}
\end{document}